\documentclass[a4paper,11pt]{article}
\usepackage{jheppub}
\RequirePackage[numbers,sort&compress]{natbib}
\usepackage{axodraw4j}
\usepackage{graphicx}
\usepackage{epsfig}
\usepackage{amssymb}
\usepackage{amsmath}
\usepackage{multirow}
\usepackage{bigdelim}
\usepackage{slashed}
\usepackage{enumerate}
\usepackage{feynarts}

\newcommand{\bea}{\begin{eqnarray}}
\newcommand{\eea}{\end{eqnarray}}
\newcommand{\bean}{\begin{eqnarray*}}
\newcommand{\eean}{\end{eqnarray*}}

\def\braket#1{\left\langle #1 \right\rangle}

\def\gb #1{ \left\langle #1 \right]}

\def\vev{\braket}

\def\bvev#1{\left[ #1 \right]}
\def\Spaa{\vev}
\def\Spbb{\bvev}
\def\Spab{\gb}

\def\Label#1{\label{#1}%
  \smash{\hbox to0pt{\raise1ex\hbox{\tiny[#1]}\hss}}}

\allowdisplaybreaks

\newcommand{\unipd}{Dipartimento di Fisica ed Astronomia, Universit\`a di Padova, Via Marzolo 8, 35131 Padova, Italy}
\newcommand{\pdinfn}{INFN, Sezione di Padova, Via Marzolo 8, 35131 Padova, Italy}
\newcommand{\mpi}{Max-Planck-Institut f\"ur Physik, F\"ohringer Ring 6, 80805 M\"unchen, Germany}
\newcommand{\SISSA}{SISSA, Via Bonomea 265, 34136 Trieste, Italy}

\title{On the Subleading-Soft Behaviour of QCD Amplitudes}
\author[a,b,c]{Hui Luo,}
\author[a,b,d]{Pierpaolo Mastrolia,}
\author[a,b]{and William J. Torres Bobadilla}

\affiliation[a]{\unipd}
\affiliation[b]{\pdinfn}
\affiliation[c]{\SISSA}
\affiliation[d]{\mpi}

\date{\today}

\abstract{We elaborate on the radiative behaviour of tree-level scattering
amplitudes in the soft regime.
We show that the sub-leading soft term in single-gluon emission of
quark-gluon amplitudes in
QCD is controlled by differential operators, whose universal form can
be derived from both on-shell recursion relation and gauge invariance, as it
was shown to hold for graviton-
and gluon-scattering.
}
\keywords{QCD, scattering amplitudes}

\begin{document}
\maketitle
\clearpage

\section{Introduction}

Soft emission of massless gauge bosons is a source of infrared singularities for scattering amplitudes. The behaviour of a radiative cross-section in the soft regime, when the momentum of the emitted particle becomes evanescent, is governed by the non-radiative process, {\it perturbed} by the action of operators that  
depend on the quantum numbers of the emitter, whose universal form is dictated by gauge invariance. 

At the tree-level, for photon-, or similarly gluon-emission, as well as for graviton-emission, the leading term in the momentum-expansion of a radiative amplitude are controlled by {\it soft factors} whose shape was identified a long ago \cite{Low:1958sn,Weinberg:1964ew}. A recent study of Cachazo and Strominger about graviton amplitudes~\cite{Cachazo:2014fwa}, followed by a similar analysis applied to gluon-scattering by Casali~\cite{Casali:2014xpa}, pointed to the existence of {\it differential operators} that control the subleading behaviour. Within the spinor formalism, the form of these operators can be easily derived by taking the soft-limit of the on-shell recursive construction of the amplitude \cite{Britto:2004ap,Britto:2005fq}. Alternatively, 
by following the proof of Low's theorem \cite{Low:1958sn}, it is possible to show that the subleading-soft operators depend on the total angular momentum (orbital and spin) of the radiator, as explicitly shown for gluon- and graviton-amplitudes by Bern, Davies, Di Vecchia, and Nohle \cite{Bern:2014vva}.
The universality character of the subleading soft operators, also in the case of fermionic emitter, was shown by White \cite{White:2014qia}, employing a set of effective Feynman rules \cite{Laenen:2010uz}.

All these recent results, in accordance to Low's theorem show that subleading soft terms are controlled by the total angular momentum of the emitter and arise from internal- and external-line emissions. While both carry informations on the orbital momentum, only the latter contain information on the spin.

The emergence of subleading soft theorems was confirmed to hold in arbitrary dimensions for tree-amplitudes of gluon and gravitons \cite{Schwab:2014xua,Afkhami-Jeddi:2014fia}, and the study of their modification due higher-order corrections has been carried out in \cite{Bern:2014oka,He:2014bga}.
The recent interest on the study of the low-energy behaviour of scattering amplitudes has been stimulating 
further investigation on how the universality of the soft expansion can be connected to additional symmetries of the S-matrix \cite{Larkoski:2014hta,Lysov:2014csa,He:2014cra,Broedel:2014fsa}. \\

In this article, we elaborate on the soft behaviour of single-gluon radiation from QCD amplitudes of two quarks and gluons.
Upon colour decomposition \cite{Giele:1991vf}, one can identify two situations according to the position of the soft gluon in the colourless ordered subamplitude: 
{\it i)} between a (anti-)quark and a gluon, and {\it ii)} between two gluons.
Case {\it (ii)} is similar to the pure Yang-Mills (YM) case, where the emitter is necessarily a gluon, and it can be considered well 
studied~\cite{Casali:2014xpa,Bern:2014vva}. 
In case {\it (i)}, instead, the soft gluon can be radiated either from a gluon or from a (anti-)quark. 
In order to derive the soft behaviour from fermion emitters, we analyse the paradigmatic case of photon bremsstrahlung from the quark-line in QED. For this case, we show the equivalence of the soft operators derived from gauge invariance and from the on-shell construction. 
This result can be, then, easily extended to the quark-gluon amplitudes in QCD.

Our derivation of the soft behaviour of quark-gluon scattering can be considered complementary to the study of subleading soft behaviour of bosonic amplitudes for graviton and YM amplitudes, 
which were addressed with similar techniques~\cite{Cachazo:2014fwa,Casali:2014xpa,Bern:2014vva}. 
Making use of no effective Feynman rule, the proof herby given can be considered as an alternative to the derivation proposed in \cite{White:2011yy}. 
Differently from the the bosonic cases, where the spin operators, being written in terms of the metric tensor, can be easily disentangled from the non-radiative amplitude, the terms from the anomalous magnetic moment require a more careful treatment. 

We explicitly apply the soft operators to describe the low-energy behaviour of quark-gluon amplitudes,
emitting either a photon or a gluon, for non-trivial helicity configurations of six-parton scattering.

The paper is organised as follows.
In Section 2 we recall the derivation of the subleading soft terms for pure Yang-Mills amplitudes by means of on-shell recursive construction, and its link to Low's theorem. In Section 3, we apply both on-shell methods and gauge invariance to derive the soft behaviour for photon bremsstrahlung from quark-gluon amplitudes.
Section 4 contains the proof of the subleading soft theorem for QCD amplitudes.
The last two sections are both accompanied by explicit examples. The calculations have been performed by using the Mathematica package S@M \cite{Maitre:2007jq}, implementing the spinor formalism.

\section{Soft limit of gluon-amplitudes}
\label{sec:2}
\begin{figure}[htb]
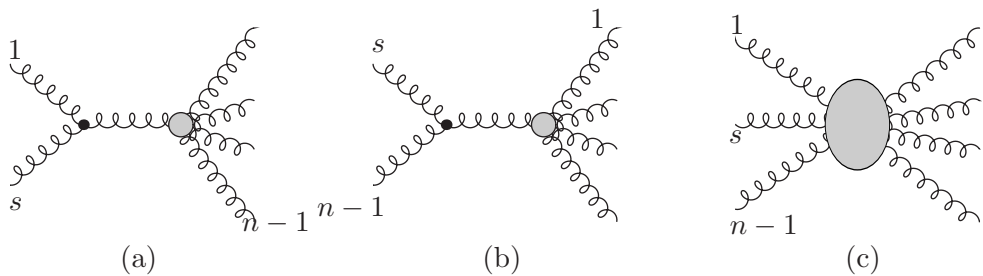

\begin{align*}
\parbox{40mm}{\input{FeynmanDiagrams/fig1a.tex}}\qquad
\parbox{40mm}{\input{FeynmanDiagrams/fig1b.tex}}\qquad
\parbox{40mm}{\input{FeynmanDiagrams/fig1c.tex}}
\end{align*}
\caption{Soft-gluon behaviour of pure-gluon amplitudes}
\label{fig:pure-gluon}
\end{figure}

According to little group transformation, the scaling behaviour of
tree-level scattering amplitudes involving massless particles is given by,
\bea
{\mathcal A}(\{t\lambda_i,t^{-1}\widetilde\lambda_i, h_i\})=t^{-2h_i}{\mathcal A}(\{\lambda_i,\widetilde\lambda_i, h_i\}),
\eea
where $h_i$ is the helicity of particle $i$,
and $\lambda_i$ and $\tilde{\lambda}_i$ are the (holomorphic and
anti-holomorphic) spinors associated to its massless momentum 
$k_{i\alpha\dot\alpha}=\lambda_{i\alpha}\widetilde \lambda_{i \dot a}$.
Let us consider the $n$-gluon amplitude ${\mathcal A}_n$, with the
gluon $s$ as soft, and momentum $k_s$ defined 
as $k_{s\alpha\dot\alpha}=\epsilon\lambda_{s\alpha}\widetilde
\lambda_{s \dot a}$, where $\epsilon$ parametrise the energy loss. The
corresponding amplitude transforms as
\bea
{\mathcal A}_n(\{\sqrt\epsilon\lambda_s,\sqrt\epsilon\widetilde\lambda_s, +1\})=\epsilon\, {\mathcal A}_n(\{\epsilon\lambda_s,\widetilde\lambda_s, +1\}).
\eea
With this choice, the soft limit is realised through the holomorphic limit
$\epsilon\lambda_s\rightarrow 0$, while keeping the anti-holomorphic
one $\widetilde \lambda_s$ as finite
\cite{Cachazo:2014fwa,Casali:2014xpa}. 

The low-energy behaviour derived from the BCFW construction \cite{Britto:2004ap,Britto:2005fq}
shows that the colour-ordered amplitude in the soft-limit reads \cite{Casali:2014xpa},
\begin{align}
{\mathcal A}_n\left(\left\{ \epsilon\,\left|s\right\rangle ,\left|s\right]\right\} ,\left\{ \left|1\right\rangle ,\left|1\right]\right\} ,\ldots,\left\{ \left|n-1\right\rangle ,\left|n-1\right]\right\} \right) & =\left(\frac{1}{\epsilon^{2}}S_{\rm G}^{\left(0\right)\lambda}+\frac{1}{\epsilon}S_{\rm G}^{\left(1\right)\lambda}\right){\mathcal A}_{n}\left(1,\ldots,n-1\right)+\mathcal{O}\left(\epsilon^{0}\right)\label{eq:SL}
\end{align}
were the leading and the sub-leading terms are,
\begin{align}
S_{\rm G}^{\left(0\right)\lambda} &
=\frac{\Spaa{n-1,\,1}}{\Spaa{s,\,1}\, \Spaa{n-1,\,s} } 
\ , \label{eq:Casali_S0}  \\
S_{\rm G}^{\left(1\right)\lambda} &
=\frac{1}{\Spaa{s,\,1}}\widetilde\lambda_s^{\dot
  \alpha}{\partial\over\partial \widetilde\lambda_1^{\dot
    \alpha}}+\frac{1}{\Spaa{n-1,\,s}}\widetilde\lambda_s^{\dot
  \alpha}{\partial\over\partial \widetilde\lambda_{n-1}^{\dot
    \alpha}} \ . \label{eq:Casali_S1}
\end{align}

As shown in \cite{Bern:2014vva}, on-shell gauge invariance can be used
as well to determine the next-to-leading soft behaviour of non-Abelian gauge theory.
The color-ordered (polarisation-stripped) amplitude
takes contribution from the three types of diagrams shown in
fig.~\ref{fig:pure-gluon}. 
Diagrams (a) and (b) contribute to the leading pole term in the soft
regime, while the third structure, with the soft-gluon emitted from
an internal propagator, is regular in this limit.
By using the on-shell gauge invariance, one obtain the soft behaviour,
\bea
A_n(k_s;k_1,\dots,k_{n-1})= \left[S_{\rm G}^{(0)}+S_{\rm
    G}^{(1)}\right] A_{n-1} (k_1,\dots,k_{n-1}) +{\cal O} (k_s) \ , 
\eea  
with 
\bea
S_{\rm G}^{(0)}&\equiv& {k_1\cdot \varepsilon(k_s;r_s)\over
  \sqrt{2}(k_1\cdot k_s)}-{k_{n-1}\cdot \varepsilon(k_s;r_s)\over
  \sqrt{2}(k_{n-1}\cdot k_s)}\label{eq:Bern_S0} \ , \\
S_{\rm G}^{(1)}&\equiv& -i\varepsilon_{\mu}(k_s;r_s) k_{s\sigma}
\left({J_{{\rm G}1}^{\mu\sigma}\over \sqrt{2}(k_1\cdot k_s)}-{J_{{\rm 
        G}n-1}^{\mu\sigma}\over \sqrt{2}(k_{n-1}\cdot
    k_s)}\right)\label{eq:Bern_S1} \ ,
\eea
where $J$ is the total angular momentum of the emitter, written in terms of the
orbital momentum $L$ and spin $\Sigma$, 
\bea
J_{{\rm G}i}^{\mu\sigma}&\equiv& L_{{\rm G}i}^{\mu\sigma}+\Sigma_{{\rm
    G}i}^{\mu\sigma} \ ,\nonumber\\
L_{{\rm G}i}^{\mu\sigma}&\equiv& i \left(k_i^\mu {\partial
    \over \partial k_{i\sigma}}-k_i^\sigma {\partial 
\over \partial k_{i\mu}} \right) \ , \nonumber\\
\Sigma_{{\rm G}i}^{\mu\sigma}&\equiv& i \left(\varepsilon_i^\mu 
  {\partial \over \partial \varepsilon_{i\sigma}}-\varepsilon_i^\sigma 
  {\partial \over \partial \varepsilon_{i\mu}} \right) \ .
\label{Gauge_boson_Jmn}
\eea
In the derivation of this result \cite{Bern:2014vva},
$L_{{\rm G}i}^{\mu\sigma}$ is understood not to act on explicit polarisation vectors, i.e. $L_{{\rm G}i}^{\mu\sigma} \varepsilon_i^\nu=0$.


As discussed in \cite{Bern:2014vva}, the equivalence between the
operators $S_{\rm  G}^{(1)}$, derived from gauge invariance, and
$S_{\rm G}^{(1)\lambda}$, derived from on-shell recurrence, can be
seen through their explicit action on polarisation vectors.
In fact,
the next-to-soft operators $S_{\rm G}^{(1)}$ and $S_{\rm G}^{(1)\lambda}$ acting on $\varepsilon^{\text{\ensuremath{\pm\rho}}}\left(k_{1};r_{1}\right)$ 
with reference momentum $k_{r_1}$ (similarly for
$\varepsilon^{\text{\ensuremath{\pm\rho}}}\left(k_{n-1};r_{n-1}\right)$) amount to,
\begin{align}
S_{\rm G}^{(1)}\varepsilon^{+\rho}(k_1;r_1)&=-{\Spaa{r_1,\, s}\over \Spaa{r_1,\, 1}\Spaa{1,\, s}}\varepsilon^{+\rho}(k_s;r_1)\label{S1YM_polar1p_GI},\\
S_{\rm G}^{(1)}\varepsilon^{-\rho}(k_1;r_1)&=+{\Spbb{r_1,\, s}\over
  \Spbb{r_1,\, 1}\Spbb{1,\,
    s}}\varepsilon^{+\rho}(k_s;r_1)\label{S1YM_polar1m_GI} \ , 
\end{align}
and
\begin{align}
S_{\rm G}^{(1)\lambda}\varepsilon^{+\rho}(k_1;r_1)&=-{\Spaa{r_1,\, s}\over \Spaa{r_1,\, 1}\Spaa{1,\, s}}\varepsilon^{+\rho}(k_s;r_1)\label{S1YM_polar1p_BCFW},\\
S_{\rm G}^{(1)\lambda}\varepsilon^{-\rho}(k_1;r_1)&=+{\Spbb{r_1,\, s}\over \Spbb{r_1,\, 1}\Spbb{1,\, s}}\left[\varepsilon^{+\rho}(k_s;r_1)-{\sqrt 2\Spbb{r_1,\, s}\over \Spbb{r_1,\, 1}\Spaa{1,\, s}}k_1^\rho\right]\label{S1YM_polar1m_BCFW}.
\end{align}
The second term in the last equation is proportional to $k_1^\rho$, and vanishes after
contracting it with the polarisation stripped $(n-1)$-point amplitude,
due to Ward identity. Therefore, the next-to-leading soft
(differential) operators obtained in the two framework are completely equivalent.

\section{Photon bremsstrahlung from quark-gluon amplitudes}
\label{sec:3}
Before turning our discussion to the soft gluon emission from
quark-gluon amplitudes,
we derive the low-energy behaviour of photon emission from colour
ordered quark-gluon tree-level amplitudes. 
Since the gluon radiation from gluon emitter has been
studied \cite{Casali:2014xpa,Bern:2014vva}, one needs to consider only the radiation
from the fermion-line. In order to isolate only this contribution,
instead of considering the gluon emission,
we consider the radiation of a photon from the quark-line of a quark-gluon amplitude.

We discuss the derivation of the leading and next-to-leading soft terms from both on-shell
recurrence and gauge invariance approaches, and proof the equivalence
of the respective results.

\subsection{Derivation from on-shell recursion relation}

\begin{figure}[tb]
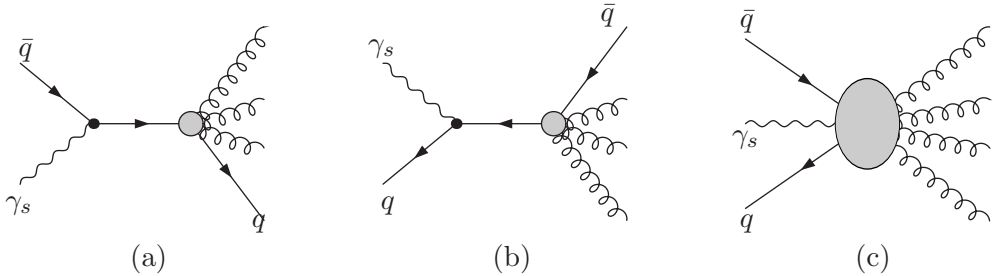

\begin{align*}
\parbox{40mm}{\input{FeynmanDiagrams/fig2a.tex}}\qquad
\parbox{40mm}{\input{FeynmanDiagrams/fig2b.tex}}\qquad
\parbox{40mm}{\input{FeynmanDiagrams/fig2c.tex}}
\end{align*}
\caption[a]{\small  Soft-photon behaviour of quark-gluon amplitudes
}\label{GluonSoftFigure}
\end{figure}
We consider a color-ordered amplitude $A_{n+3}(\Lambda_{\bar q},
\gamma_s^+,\Lambda_q, g_1,\cdots, g_n)$, 
where $\Lambda_q$ and $\Lambda_{\bar q}$ denote the quark and
antiquark, and $\gamma_s^+$ stands for a soft photon of helicity $h_s=+1$, emitted
from the fermionic current. The explicit helicity choice of the photon
does not affect the generality of our derivation, since the
negative-helicity case can be treated analogously. 
Under the BCFW deformation involving $\gamma_s$ and $g_n$ pair \cite{Luo:2005my}
\bea
&&\left|\widehat{s}\right\rangle =\left|s\right\rangle
+z\,\left|n\right\rangle \ , \label{eq:Sn-1}\\
&&\left|\widehat{n}\right]=\left|n\right]-z\,\left|s\right] \ , \label{eq:Sn}
\eea
the amplitude factorises as,
\begin{align}
A_{n+3}\left(\Lambda_{\bar{q}},\gamma_{s}^{+},\Lambda_{q},g_{1},\cdots,g_{n}\right) & =\sum A_{L}\left(\widehat{\gamma_{s}^{+}},\Lambda_{q},\widehat{I}\right)\frac{1}{P_{s,q}^{2}}A_{R}\left(-\widehat{I},g_{1},\ldots,\widehat{g_{n}},\Lambda_{\bar{q}}\right)\nonumber \\
 & \qquad+\sum A_{L}\left(\Lambda_{\bar{q}},\widehat{\gamma_{s}^{+}},\widehat{I}\right)\frac{1}{P_{\bar{q},s}^{2}}A_{R}\left(-\widehat{I},\Lambda_{q},g_{1}\ldots,\widehat{g_{n}}\right)\nonumber \\
 & \qquad+\sum
 A_{L}\left(\Lambda_{\bar{q}},\widehat{\gamma_{s}^{+}},\Lambda_{q},g_{1},\ldots,g_{j},\widehat{I}\right)\frac{1}{P_{\bar{q},s,q,1,\dots,j}^{2}}A_{R}\left(-\widehat{I},g_{j+1},\ldots,\widehat{g_{n}}\right)
 \ ,
\label{eq:bcfw:qqg}
\end{align}
as depicted in fig.~\ref{GluonSoftFigure}. 
In the first term, on-shellness requires $z=-{\Spaa{q,s}\over\Spaa{q,n}}$, and
$\widehat I$ stands for a fermion with opposite helicity with respect
to $q$.  By taking the soft limit, $|s\rangle \rightarrow \epsilon
|s\rangle$, this terms reads
\begin{multline}
\frac{1}{\epsilon^{2}}\frac{\Spaa{n,q}}{\Spaa{n,s}\Spaa{s,q}}\times A_{n+2}\left(\{|q\rangle,|q]+\epsilon\frac{\Spaa{n,s}}{\Spaa{n,q}}|s]\},\{|1\rangle,|1]\},\ldots,\{|n\rangle,|n]+\epsilon\frac{\Spaa{s,q}}{\Spaa{n,q}}|s]\},\{|\bar{q}\rangle,|\bar{q}]\}\right)\\
=\left[\frac{1}{\epsilon^{2}}\frac{\Spaa{n,q}}{\Spaa{n,s}\Spaa{s,q}}+\frac{1}{\epsilon}\left(\frac{1}{\Spaa{s,q}}\widetilde{\lambda}_{s}^{\dot{a}}\frac{\partial}{\partial\widetilde{\lambda}_{q}^{\dot{a}}}+\frac{1}{\Spaa{n,s}}\widetilde{\lambda}_{s}^{\dot{a}}\frac{\partial}{\partial\widetilde{\lambda}_{n}^{\dot{a}}}\right)\right]A_{n+2}\left(\{|q\rangle,|q]\},\{|1\rangle,|1]\},\ldots,\{|n\rangle,|n]\},\{|\bar{q}\rangle,|\bar{q}]\}\right).
\end{multline}
In the second term of (\ref{eq:bcfw:qqg}), the on-shell condition implies $z=-{\Spaa{\bar q,s}\over\Spaa{\bar
    q,n}}$, and $\widehat I$ stands for a fermion with opposite
helicity with respect to $\bar q$. In the soft limit, $|s\rangle
\rightarrow \epsilon |s\rangle$, this term becomes
\begin{multline}
-\frac{1}{\epsilon^{2}}\frac{\Spaa{n,\bar{q}}}{\Spaa{n,s}\Spaa{s,\bar{q}}}\times A_{n+2}\left(\{|\bar{q}\rangle,|\bar{q}]+\epsilon\frac{\Spaa{n,s}}{\Spaa{n,\bar{q}}}|s]\},\{|q\rangle,|q]\},\{|1\rangle,|1]\},\ldots,\{|n\rangle,|n]+\epsilon\frac{\Spaa{s,q}}{\Spaa{n,q}}|s]\}\right)\\
=-\left[\frac{1}{\epsilon^{2}}\frac{\Spaa{n,\bar{q}}}{\Spaa{n,s}\Spaa{s,\bar{q}}}+\frac{1}{\epsilon}\left(\frac{1}{\Spaa{s,\bar{q}}}\widetilde{\lambda}_{s}^{\dot{a}}\frac{\partial}{\partial\widetilde{\lambda}_{\bar{q}}^{\dot{a}}}+\frac{1}{\Spaa{n,s}}\widetilde{\lambda}_{s}^{\dot{a}}\frac{\partial}{\partial\widetilde{\lambda}_{n}^{\dot{a}}}\right)\right]A_{n+2}\left(\{|\bar{q}\rangle,|\bar{q}]\},\{|q\rangle,|q]\},\{|1\rangle,|1]\},\ldots,\{|n\rangle,|n]\}\right).
\end{multline}
As shown in Appendix \ref{sec:Appendix}, the third term of (\ref{eq:bcfw:qqg}) contains only finite
contribution, hence does not contribute to any soft operator.
After summing the first and second term together, the amplitude in
the soft regime reads,
\bea
A_{n+3}\left(\Lambda_{\bar q},\gamma_s^+,\Lambda_q, g_1,\cdots,
  g_n\right)=\left({1\over \epsilon^2}S^{(0)\lambda}+{1\over
    \epsilon}S^{(1)\lambda}\right)A_{n+2}\left(\Lambda_{\bar
    q},\Lambda_q, g_1,\cdots, g_n\right) + {\cal O}(1) \ , \label{NS_QCD_BCFW}
\eea
with
\bea
S^{(0)\lambda}&=&{\Spaa{n,q}\over \Spaa{n,s}\Spaa{s,q}}- {\Spaa{n,\bar q}\over \Spaa{n,s}\Spaa{s,\bar q}}={\Spaa{\bar q,\, q}\over  \Spaa{\bar q,\,s} \Spaa{s,\,q} },\label{S0_f_BCFW}\\
S^{(1)\lambda}&=&{1\over \Spaa{s, q}}\widetilde \lambda_s^{\dot a}{\partial\over \partial \widetilde \lambda_{q}^{\dot a}}- {1\over \Spaa{s, \bar q}}\widetilde \lambda_s^{\dot a}{\partial\over \partial \widetilde \lambda_{\bar q}^{\dot a}},\label{S1_f_BCFW}
\eea 
and where $A_{n+2}$ is the non radiative quark-gluon amplitude.

\subsection{Derivation from gauge invariance}
\label{subsec:QCD-gauge-inv}

Within a diagrammatic approach, 
the amplitude $A\left(\Lambda_{\bar q},\gamma_s^+,\Lambda_q,
  g_1,\cdots, g_n\right)$ 
gets contribution from the three diagrams in
fig.~\ref{GluonSoftFigure}, and can be obtained by contracting the
soft-photon polarisation $\varepsilon_\mu(k_s;r_s)$ and the current
$A_{n+3}^{\mu} $, 
\bea
A_{n+3} = \varepsilon_\mu(k_s;r_s) \ A_{n+3}^{\mu} \ ,
\eea
with 
\begin{align}
A_{n+3}^{\mu}\left(k_{s};k_{\bar{q}},k_{q},k_{1},\dots,k_{n}\right) & =-\frac{i}{\sqrt{2}}{\overline{u}}(k_{q})\tilde{A}(k_{\bar{q}}+k_{s},\, k_{q},\, k_{1},\dots,k_{n})\frac{i\left(\slashed{k}_{\bar{q}}+\slashed{k}_{s}\right)}{(k_{\bar{q}}+k_{s})^{2}}\gamma^{\mu}v(k_{\bar{q}})\nonumber \\
 &
 \quad+\frac{i}{\sqrt{2}}{\overline{u}}(k_{q})\gamma^{\mu}\frac{i\left(\slashed{k}_{q}+\slashed{k}_{s}\right)}{(k_{q}+k_{s})^{2}}\tilde{A}(k_{\bar{q}},\,
 k_{q}+k_{s},\, k_{1},\dots,k_{n})v(k_{\bar{q}}) \nonumber \\
 & \quad+N_{n+3}^{\mu}(k_{s};k_{\bar{q}},k_{q},k_{1},\dots,k_{n}) \ . 
\end{align}
The first term, corresponding to fig.~\ref{GluonSoftFigure} (a), represents the case of soft-photon emission from an outgoing antiquark;
the second term, corresponding to fig.~\ref{GluonSoftFigure} (a),
represents the case of soft-photon emission from an outgoing quark;
while the third term
$N_{n+3}^{\mu}$ represents
the case of soft-photon emission from internal fermion lines, as shown in fig.~\ref{GluonSoftFigure} (c).
  $\widetilde A$ is the internal part sandwiched by two free-particle states of fermions.
By using the anti-commuting $\gamma$-matrix relations, the massless Dirac equation,
$\overline u(p)\,\slashed p=\slashed p \,v(p)=0$, 
the transversality conditions, $k_{s}\cdot\varepsilon^{\pm}(k_s;r_s)=0$
and the relation 
$
\gamma^{\mu}\slashed p 
=p_{\nu} (\eta^{\mu\nu}+[\gamma^{\mu},\gamma^{\nu}]/2)
$
(for an arbitrary momentum $p$),
the current $A_{n+3}^{\mu}$ can be cast as,
\begin{align}
 A_{n+3}^{\mu}\left(k_{s};k_{\bar{q}},k_{q},k_{1},\dots,k_{n}\right)
&=\frac{k_{\bar{q}}^{\mu}}{\sqrt{2}k_{\bar{q}}\cdot k_{s}}{\overline{u}}(k_{q})\tilde{A}(k_{\bar{q}}+k_{s},\, k_{q},\, k_{1},\dots,k_{n})v(k_{\bar{q}})\nonumber\\%
&+\frac{i\,{k}_{s\nu}}{\sqrt{2}k_{\bar{q}}\cdot k_{s}}{\overline{u}}(k_{q})\tilde{A}(k_{\bar{q}}+k_{s},\, k_{q},\, k_{1},\dots,k_{n})\Sigma_{F}^{\mu\nu}v(k_{\bar{q}})\nonumber\\%
&-\frac{k_{q}^{\mu}}{\sqrt{2}k_{q}\cdot k_{s}}{\overline{u}}(k_{q})\tilde{A}(k_{\bar{q}},\, k_{q}+k_{s},\, k_{1},\dots,k_{n})v(k_{\bar{q}})\nonumber\\%
&+\frac{i\,{k}_{s\nu}}{\sqrt{2}k_{q}\cdot k_{s}}{\overline{u}}(k_{q})\,\Sigma_{F}^{\mu\nu}\,\tilde{A}(k_{\bar{q}},\, k_{q}+k_{s},\, k_{1},\dots,k_{n})v(k_{\bar{q}})\nonumber\\%
&+N_{n+3}^{\mu}(k_{s};k_{\bar{q}},k_{q},k_{1},\dots,k_{n})\ , 
\label{GI_transform}
\end{align}
where 
\bea
\Sigma_F^{\mu \nu}\equiv {i\over 4}\left[\gamma^\mu,\gamma^\nu\right]\label{Sigma_fermion}
\eea
is the spin operator in a 4-dimensional representation of the Lorentz algebra corresponding to spin $1/2$.
Following \cite{Low:1958sn,Bern:2014vva}, we can determine 
$N^\mu$ by imposing gauge invariance. In fact, the condition 
\bea
k_{s\,\mu}\,A_{n+3}^\mu (k_s; k_{\bar q}, k_q, k_1, \dots, k_n)=0 \ ,
\eea 
together with on-shell massless condition $k_s^2=0$, implies 
\begin{multline}
k_{s\,\mu}N_{n+3}^{\mu}(0;k_{q},k_{\bar{q}},k_{1},\dots,k_{n})\\
=-\frac{k_{s\,\mu}}{\sqrt{2}}{\overline{u}}(k_{q})\left[\frac{\partial}{\partial
    k_{\bar{q}\mu}}\tilde{A}(k_{\bar{q}},\, k_{q},\,
  k_{1},\dots,k_{n})-\frac{\partial}{\partial
    k_{q\mu}}\tilde{A}(k_{\bar{q}},\, k_{q},\,
  k_{1},\dots,k_{n})\right]v(k_{\bar{q}}) \ .
\end{multline}
Consequently, we can write $A_{n+3}^{\mu}$ as, 
\begin{align}
 & A_{n+3}^{\mu}(k_{s};k_{\bar{q}},k_{q},k_{1},\dots,k_{n})=\left(\frac{k_{\bar{q}}^{\mu}}{\sqrt{2}k_{\bar{q}}\cdot k_{s}}-\frac{k_{q}^{\mu}}{\sqrt{2}k_{q}\cdot k_{s}}\right){\overline{u}}(k_{q})\tilde{A}(k_{\bar{q}},\, k_{q},\, k_{1},\dots,k_{n})v(k_{\bar{q}})\nonumber \\
 & \qquad\qquad+\frac{i\,{k}_{s\nu}}{\sqrt{2}k_{\bar{q}}\cdot k_{s}}{\overline{u}}(k_{q})\tilde{A}(k_{\bar{q}},\, k_{q},\, k_{1},\dots,k_{n})\Sigma_{F}^{\mu\nu}v(k_{\bar{q}})\nonumber \\
 & \qquad\qquad+\frac{i\,{k}_{s\nu}}{\sqrt{2}k_{q}\cdot k_{s}}{\overline{u}}(k_{q})\,\Sigma_{F}^{\mu\nu}\, \tilde{A}(k_{\bar{q}},\, k_{q},\, k_{1},\dots,k_{n})v(k_{\bar{q}})\nonumber \\
 & \qquad\qquad-\frac{i\, k_{s\,\nu}}{\sqrt{2}k_{\bar{q}}\cdot k_{s}}{\overline{u}}(k_{q})\left[i\left(k_{\bar{q}}^{\mu}\frac{\partial}{\partial k_{\bar{q}\nu}}-k_{\bar{q}}^{\nu}\frac{\partial}{\partial k_{\bar{q}\mu}}\right)\tilde{A}(k_{\bar{q}},\, k_{q},\, k_{1},\dots,k_{n})\right]v(k_{\bar{q}})\nonumber \\
 & \qquad\qquad+\frac{i\, k_{s\,\nu}}{\sqrt{2}k_{q}\cdot k_{s}}{\overline{u}}(k_{q})\left[i\left(k_{q}^{\mu}\frac{\partial}{\partial k_{q\nu}}-k_{q}^{\nu}\frac{\partial}{\partial k_{q\mu}}\right)\tilde{A}(k_{\bar{q}},\, k_{q},\, k_{1},\dots,k_{n})\right]v(k_{\bar{q}})\nonumber\\
 &\qquad\qquad+\mathcal{O}\left(k_{s}\right)\, .
\end{align}
Then, we contract back the polarisation vector of the soft-photon $\varepsilon^+_\mu(k_s;r_s)$, obtaining the following expression of the amplitude, 
\begin{align}
A_{n+3}(k_{s};k_{\bar{q}},k_{q},k_{1},\dots,k_{n})= & \left(\frac{\varepsilon^{+}(k_s;r_s)\cdot k_{\bar{q}}}{\sqrt{2}k_{\bar{q}}\cdot k_{s}}-\frac{\varepsilon^{+}(k_s;r_s)\cdot k_{q}}{\sqrt{2}k_{q}\cdot k_{s}}\right)A_{n+2}(k_{s};k_{\bar{q}},k_{q},k_{1},\dots,k_{n})\nonumber \\
 & \quad+\frac{i\,\varepsilon_{\mu}^{+}(k_s;r_s)\,{k}_{s\nu}}{\sqrt{2}k_{\bar{q}}\cdot k_{s}}{\overline{u}}(k_{q})\tilde{A}(k_{\bar{q}},\, k_{q},\, k_{1},\dots,k_{n})\Sigma_{F}^{\mu\nu}v(k_{\bar{q}})\nonumber \\
 & \quad+\frac{i\,\varepsilon_{\mu}^{+}(k_s;r_s)\,{k}_{s\nu}}{\sqrt{2}k_{q}\cdot k_{s}}{\overline{u}}(k_{q})\,\Sigma_{F}^{\mu\nu}\, \tilde{A}(k_{\bar{q}},\, k_{q},\, k_{1},\dots,k_{n})v(k_{\bar{q}})\nonumber \\
 & \quad-\frac{i\,\varepsilon_{\mu}^{+}(k_s;r_s)\, k_{s\,\nu}}{\sqrt{2}}{\overline{u}}(k_{q})\left[\left(\frac{L_{\bar{q}}^{\mu\nu}}{k_{\bar{q}}\cdot k_{s}}-\frac{L_{q}^{\mu\nu}}{k_{q}\cdot k_{s}}\right)\tilde{A}(k_{\bar{q}},\, k_{q},\, k_{1},\dots,k_{n})\right]v(k_{\bar{q}})\nonumber\\
 &\quad+\mathcal{O}\left(k_{s}\right)\, ,
\label{totampli0}
\end{align}
with $A_{n+2}(k_s; k_{\bar q}, k_q, k_1, \dots, k_n)={\overline
  u}(k_q) \tilde{A}(k_{\bar q},\,k_q,\,k_1, \dots, k_n) v(k_{\bar q})$
being the lower-point,
non-radiative amplitude.
In the above expression, 
\bea
L^{\mu\nu}_{f_i}= i\left(k_i^\mu{\partial\over \partial k_{i
      \nu}}-k_i^\nu{\partial\over \partial k_{i \mu}}\right) \ ,
\label{L-fermion} 
\eea
is the orbital angular momentum of fermion $i$, which does not act on Dirac fields of incoming/outgoing spin $1/2$ particles, namely
\bea
L^{\mu\nu}_{f_i} u_\pm(k_i)= L^{\mu\nu}_{f_i} v_\pm(k_i)=\overline
u_\pm(k_i) L^{\mu\nu}_{f_i} 
=\overline v_\pm(k_i) L^{\mu\nu}_{f_i}=0 \ . 
\eea
On the other hand, the Lorentz generators $\Sigma_F^{\mu\nu}$ of spin $1/2$ only act on the Dirac fields $u_\pm(k_i)$ or $v_\pm(k_i)$ ($\overline u_\pm(k_i)$ or $\overline v_\pm(k_i)$).

\subsection{Connection between the two derivations}
\label{subsec:connection_QCD}
We will momentarily proof the equivalence of the limiting behaviour of
quark-gluon amplitudes in the soft-photon emission regime obtained in
(\ref{NS_QCD_BCFW}) 
from BCFW recurrence, and in (\ref{totampli0})  from gauge invariance. 

\paragraph {A. Leading soft singularity}\hspace{0pt} \\
The leading singularity factor of order $1/ k_s$ in (\ref{totampli0}) can be denoted as 
\bea
S^{(0)}={\varepsilon^+(k_s;r_s)\cdot k_{\bar q}\over \sqrt 2 k_{\bar q}\cdot k_s}-{\varepsilon^+(k_s;r_s)\cdot  k_q \over \sqrt 2 k_q\cdot k_s}.
\eea
In spinorial notations, we indeed find,
\bea
S^{(0)}={\varepsilon^+(k_s;r_s)\cdot k_{\bar q}\over \sqrt 2 k_{\bar
    q}\cdot k_s}-{\varepsilon^+(k_s;r_s)\cdot  k_q \over \sqrt 2
  k_q\cdot k_s}={\Spaa{ q,\,\bar q}\over  \Spaa{\bar q,\,s}
  \Spaa{s,\,q} }=\,-S^{(0)\lambda} \ , 
\label{S0-QED-equivalence}
\eea
where $S^{(0)\lambda}$ was defined (\ref{S0_f_BCFW}).
This prove the equivalence of the leading soft term in the two
approaches (up to an overall sign)

\paragraph {B. Next-to-leading  soft singularity}\hspace{0pt} \\
The action of the differential operator $S^{(1)\lambda}$, defined in
(\ref{S1_f_BCFW}), on the lower-point amplitude 
$A_{n+2}(\bar q,\,q,\,g_1\,\dots,\, g_n)$ can be written as,
\begin{align}
S^{(1)\lambda}A_{n+2}(\bar{q},\, q,\, g_{1}\,\dots,\, g_{n}) &
=S^{(1)\lambda}\left[\overline{u}(k_{q})\tilde{A}(k_{\bar{q}},\,
  k_{q},\, k_{1},\,\dots,\, k_{n})\, v(k_{\bar{q}})\right]  \nonumber \\
 & =\overline{u}(k_{q})\tilde{A}(k_{\bar{q}},\, k_{q},\, k_{1},\,\dots,\, k_{n})\,\left[S^{(1)\lambda}\, v(k_{\bar{q}})\right]\nonumber \\
 & \qquad+\left[S^{(1)\lambda}\overline{u}(k_{q})\right]\tilde{A}(k_{\bar{q}},\, k_{q},\, k_{1},\,\dots,\, k_{n})\, v(k_{\bar{q}})\nonumber \\
 & \qquad+\overline{u}(k_{q})\left[S^{(1)\lambda}\, \tilde{A}(k_{\bar{q}},\, k_{q},\, k_{1},\,\dots,\, k_{n})\right]\, v(k_{\bar{q}})\, .
\label{NS_BCFW_compare}
\end{align}
On the other hand, the next-to-leading soft singularity as derived in sec. \ref{subsec:QCD-gauge-inv} is
\begin{align}
A_{n+3}(k_{s};k_{\bar{q}},k_{q},k_{1},\dots,k_{n})\big|_{S^{(1)}}&= \frac{i\, \varepsilon_{\mu}^{+}(k_s;r_s)\,{k}_{s\nu}}{\sqrt{2}k_{\bar{q}}\cdot k_{s}}{\overline{u}}(k_{q})\tilde{A}(k_{\bar{q}},\, k_{q},\, k_{1},\dots,k_{n})\Sigma_{F}^{\mu\nu}v(k_{\bar{q}})\nonumber \\
 & \quad+\frac{i\, \varepsilon_{\mu}^{+}(k_s;r_s)\,{k}_{s\nu}}{\sqrt{2}k_{q}\cdot k_{s}}{\overline{u}}(k_{q})\,\Sigma_{F}^{\mu\nu}\, \tilde{A}(k_{\bar{q}},\, k_{q},\, k_{1},\dots,k_{n})v(k_{\bar{q}})\nonumber \\
 & \quad-i\frac{\varepsilon_{\mu}^{+}(k_s;r_s)\, k_{s\,\nu}}{\sqrt{2}}{\overline{u}}(k_{q})\left[\left(\frac{L_{\bar{q}}^{\mu\nu}}{k_{\bar{q}}\cdot k_{s}}-\frac{L_{q}^{\mu\nu}}{k_{q}\cdot k_{s}}\right)\tilde{A}(k_{\bar{q}},\, k_{q},\, k_{1},\dots,k_{n})\right]v(k_{\bar{q}})\, .
\label{NS_GI_compare}
\end{align}
We proceed by identifying (\ref{NS_BCFW_compare}) and (\ref{NS_GI_compare}) term by term. 
\begin{itemize}

\item
{\bf Proposition 1:}
\bea
S^{(1)\lambda}\,v(k_{\bar q})= -\left[{i\,\,\varepsilon^+_\mu(k_s;r_s)\, {k}_{s\nu} \over \sqrt 2 k_{\bar q}\cdot k_s}\,\Sigma^{\mu \nu}\,v(k_{\bar q})\right]\, .
\eea
{\it proof.}

Outgoing antiquark with different helicities in terms of spinor notations are
\begin{align}
h_{\bar q}=+{1\over 2},&\qquad\,v_+(k_{\bar q})=\widetilde \lambda_{\bar q}^{\dot \alpha}=|\bar q]\, ,\\
h_{\bar q}=-{1\over 2},&\qquad\,v_-(k_{\bar q})=(\lambda_{\bar q})_\alpha=|\bar q\rangle\, .
\end{align} 
For $h_{\bar q}=+{1\over 2}$ case, the action of next-to-soft operator on field $v_+(k_{\bar q})$ corresponding to the first term in (\ref{NS_BCFW_compare}) is
\bea
S^{(1)\lambda}\,v_+(k_{\bar q})= \left({1\over \Spaa{s, q}}\widetilde \lambda_s^{\dot a}{\partial\over \partial \widetilde \lambda_{q}^{\dot a}}- {1\over \Spaa{s, \bar q}}\widetilde \lambda_s^{\dot a}{\partial\over \partial \widetilde \lambda_{\bar q}^{\dot a}} \right)\widetilde \lambda_{\bar q}^{\dot b}=\,-\,{1\over \Spaa{s,\,\bar q}} |s]\, ,\label{S1_BCFW_qbar_hp}
\eea
and the counter-part from (\ref{NS_GI_compare}) in terms of spinor notations is
\bea
{i\,\,\varepsilon^+_{\mu}(k_s;r_s)\, {k}_{s\nu} \over \sqrt 2 k_{\bar q}\cdot k_s}\,\Sigma_F^{\mu \nu}\,v_+(k_{\bar q})
=+\,{1\over \Spaa{s,\,\bar q}}  |s]\, .
\label{S1_GI_qbar_hp}
\eea
The result of (\ref{S1_BCFW_qbar_hp}) and (\ref{S1_GI_qbar_hp}) differ
only for a minus sign.

On the other side, for $h_{\bar q}=-{1\over 2}$ case
\bea
S^{(1)\lambda}\,v_-(k_{\bar q})= {i\,\,\varepsilon^+_\mu(k_s;r_s)\, {k}_{s\nu} \over \sqrt 2 k_{\bar q}\cdot k_s}\,\Sigma^{\mu \nu}\,v_-(k_{\bar q})
=0\, .
\label{S1_GI_qbar_hm}
\eea

\item
{\bf Proposition 2:}
\bea
S^{(1)\lambda}\,\overline u(k_q)\,=-\left[\overline u(k_q) {i\,\,\varepsilon^+_{\mu}(k_s;r_s)\, {k}_{s\nu} \over \sqrt 2 k_q\cdot k_s}\,\Sigma_F^{\mu \nu}\right]\, .
\eea

{\it proof.}

Outgoing quark with different helicities in terms of spinor notations are
\begin{align}
h=+{1\over 2},&\qquad\overline u_+(k_q)=(\widetilde \lambda_q)_{\dot a} =[\bar q|\, ,\\
h=-{1\over 2},&\qquad\overline u_-(k_q)=\lambda_q^a=\langle q|\, .
\end{align}
With a similar procedure dealing with outgoing antiquark, we have 
\bea
S^{(1)\lambda}\,\overline u_+(k_q)\,=
+\,{1\over \Spaa{s,\,q}} [s| = 
-\left(\overline u_+(k_q) {i\,\,\varepsilon^+_{\mu}(k_s;r_s)\,
    {k}_{s\nu} \over \sqrt 2 k_q\cdot k_s}\,\Sigma_F^{\mu \nu}\right)
\ , 
\eea
and
\bea
S^{(1)\lambda}\, \overline u_-(k_q)\,=\overline u_-(k_q) {i\,\,\varepsilon^+_{\mu}(k_s;r_s)\, {k}_{s\nu} \over \sqrt 2 k_q\cdot k_s}\,\Sigma_F^{\mu \nu}=0\, .
\eea

\item
{\bf Proposition 3:}
\begin{align}
S^{(1)\lambda}\tilde{A}(k_{\bar{q}},k_{q},k_{1},\dots, k_{n})=
i\frac{\varepsilon_{\mu}^{+}(k_s;r_s) k_{s\nu}}{\sqrt{2}}\left[\left(\frac{L_{\bar{q}}^{\mu\nu}}{k_{\bar{q}}\cdot k_{s}}-\frac{L_{q}^{\mu\nu}}{k_{q}\cdot k_{s}}\right)\tilde{A}(k_{\bar{q}}, k_{q}, k_{1},\dots,k_{n})\right]\, .
\end{align}

{\it proof.}

$\widetilde A$ consists of gluon polarisation and momenta, of
quark and antiquark momenta, and of $\gamma$ matrices, and it 
can be expressed in terms of spinor chains.

Since tree-level amplitudes are rational functions of spinor products, we can focus on the action of the operators $S^{(1)\lambda}$ and $-i g\varepsilon^+_{\mu}(k_s;r_s)k_{s\,\nu} \left[L_{\bar q}^{\mu\nu}/(k_{\bar q}\cdot k_s) -L_q^{\mu\nu}/(k_q\cdot k_s)\right]/\sqrt 2 $ onto each ingredient separately.

The next-to-leading soft operator gives non-trivial result when acting on spinor products involving ${\overline q}$, as  
\begin{align}
S^{(1)\lambda}\, [\bullet , \bar q]= -\,{1\over \Spaa{s,\,\bar q}}  [\bullet , s]\,,
&\quad&
S^{(1)\lambda}\, {1\over [p,\bar q]}=+\,{1\over \Spbb{p,\,\bar q}} {\Spbb{p,\,s}\over \Spaa{s,\,\bar q}\Spbb{p,\,\bar q}}\, .  \label{S1B-on-Mqbar}
\end{align}
On the other hand, the operator from gauge invariance acts on terms
depending on the antiquark momentum, like 
\begin{align}
-\frac{i\,\varepsilon_{\mu}^{+}(k_s;r_s)\, k_{s\,\nu}}{\sqrt{2}}\left(\frac{L_{\bar{q}}^{\mu\nu}}{k_{\bar{q}}\cdot k_{s}}-\frac{L_{q}^{\mu\nu}}{k_{q}\cdot k_{s}}\right)k_{\bar{q}}^{\rho} & =+\frac{1}{\Spaa{s,\,\bar{q}}}\frac{\Spab{\bar{q}|\gamma^{\rho}|s}}{2}\, ,\label{S1G-on-Mqbar}\\
-\frac{i\,\varepsilon_{\mu}^{+}(k_s;r_s)\, k_{s\,\nu}}{\sqrt{2}}\left(\frac{L_{\bar{q}}^{\mu\nu}}{k_{\bar{q}}\cdot k_{s}}-\frac{L_{q}^{\mu\nu}}{k_{q}\cdot k_{s}}\right)\frac{1}{p\cdot k_{\bar{q}}} & =-\,\frac{1}{p\cdot k_{\bar{q}}}\frac{\Spbb{p,\, s}}{\Spaa{s,\,\bar{q}}\Spbb{p,\,\bar{q}}}\, .
\label{S1G-on-Mqbar_inv}
\end{align}
The coefficients are identical up to an overall minus sign. 

The same conclusion can be drawn from comparing the action of the
subleading soft operators on the variables associated to quarks,
\begin{align}
&S^{(1)\lambda}\,[\bullet, q]=+\,\frac{1}{\Spaa{s,\, q}}[\bullet, s],&&S^{(1)\lambda}\,\frac{1}{[p,\, q]}=-\,\frac{1}{\Spbb{p,\, q}}\frac{\Spbb{p,\, s}}{\Spaa{s,\, q}\Spbb{p,\, q}}\label{S1B-on-Mq}\, ,
\end{align}
and 
\begin{align}
-\frac{i\,\varepsilon_{\mu}^{+}(k_s;r_s)\, k_{s\,\nu}}{\sqrt{2}}\left(\frac{L_{\bar{q}}^{\mu\nu}}{k_{\bar{q}}\cdot k_{s}}-\frac{L_{q}^{\mu\nu}}{k_{q}\cdot k_{s}}\right)k_{q}^{\rho}&=-\frac{1}{\Spaa{s,\, q}}\frac{\Spab{q|\gamma^{\rho}|s}}{2}\, ,\label{S1G-on-Mq}\\
-\frac{i\,\varepsilon_{\mu}^{+}(k_s;r_s)\, k_{s\,\nu}}{\sqrt{2}}\left(\frac{L_{\bar{q}}^{\mu\nu}}{k_{\bar{q}}\cdot k_{s}}-\frac{L_{q}^{\mu\nu}}{k_{q}\cdot k_{s}}\right)\frac{1}{p\cdot k_{q}}&=+\frac{1}{p\cdot k_{q}}\frac{\Spbb{p,\, s}}{\Spaa{s,\, q}\Spbb{p,\, q}}\, .\label{S1G-on-Mq_inv}
\end{align}

\end{itemize}

This complete the proof that the (leading and sub-leading) soft
operators derived from BCFW recurrence and gauge invariance are indeed
equivalent (up to an overall minus sign).

\subsection{Examples}
We consider an amplitude with one quark-antiquark pair and gluons, and
a plus-helicity photon emitted from the fermion line.
\subsubsection{MHV and $\overline{\bf \rm MHV}$ amplitudes}
\begin{itemize}
\item
The MHV amplitude is 
\bea
A_{n+3}(\Lambda_{\bar q}^+,\gamma_s^+,\Lambda_q^-, g_1^+,\cdots, g_I^-,\cdots, g_n^+) = {i\Spaa{q,\, I}^3\,\Spaa{\bar q,\, I}\over \Spaa{\bar q,\,s} \Spaa{s,\,q} \cdots \Spaa{n,\,\bar q}}\, .
\eea
By taking the holomorphic soft limit $|s\rangle\rightarrow \epsilon
|s\rangle$, one gets,
\bea
A_{n+3}(\Lambda_{\bar q}^+,\gamma_s^+,\Lambda_q^-, g_1^+,\cdots, g_I^-,\cdots, g_n^+)\bigg|_{|s\rangle\rightarrow \epsilon |s\rangle} ={1\over \epsilon^2}{\Spaa{\bar q,\, q}\over  \Spaa{\bar q,\,s} \Spaa{s,\,q} }\times {i\Spaa{q,\, I}^3\,\Spaa{\bar q,\, I}\over \Spaa{\bar q,\, q} \Spaa{1,\, 2}\cdots \Spaa{n,\,\bar q}}\, .
\eea
On the other hand, the action of the operators $S^{(0)\lambda}$ and
$S^{(1)\lambda}$ on the lower-point, non-radiative amplitude, $A(\Lambda_{\bar q}^+,\Lambda_q^-, g_1^+,\cdots, g_I^-,\cdots, g_n^+) $ reads,
\begin{align}
\left(\frac{1}{\epsilon^{2}}S^{(0)\lambda}+\frac{1}{\epsilon}S^{(1)\lambda}\right)A_{n+2}(\Lambda_{\bar{q}}^{+},\Lambda_{q}^{-},g_{1}^{+},\cdots,g_{I}^{-},\cdots,g_{n}^{+}) & =\frac{1}{\epsilon^{2}}\frac{\Spaa{\bar{q},\, q}}{\Spaa{\bar{q},\, s}\Spaa{s,\, q}}\times\frac{i\Spaa{q,\, I}^{3}\,\Spaa{\bar{q},\, I}}{\Spaa{\bar{q},\, q}\Spaa{1,\,2}\cdots\Spaa{n,\,\bar{q}}}\, .
\end{align}
The two results are identical, and in particular, for the MHV
amplitude, no next-to-soft contribution arises,
because 
$S^{(1)\lambda} A_{n+2}(\Lambda_{\bar{q}}^{+},\Lambda_{q}^{-},g_{1}^{+},\cdots,g_{I}^{-},\cdots,g_{n}^{+})
=0$. 

\item
The case of the $\overline{\rm MHV}$ amplitude is trivial, because,
since the amplitude has an anti-holomorphic expression,
\bea
A_{n+3}(\Lambda_{\bar q}^+, \gamma_s^+,\Lambda_q^-, g_1^-,\cdots,\cdots, g_n^-) &=& {i\Spbb{\bar q,\,s}^3 \Spbb{q,\,s} \over\Spbb{ \bar q,\,s} \Spbb{s,\,q}\Spbb{q,\,1} \Spbb{1,\,2} \cdots \Spbb{ n,\,\bar q} }\, ,
\eea
the holomorphic soft limit $|s\rangle\rightarrow \epsilon |s\rangle$ 
has no effect.
At the same time, the soft operators $S^{(0)\lambda}$ and
$S^{(1)\lambda}$ should act on a vanishing amplitude, because the lower amplitude $A_{n+2}(\Lambda_{\bar q}^+, \Lambda_q^-, g_1^-,\cdots,\cdots, g_n^-)=0$.
\end{itemize}

\subsubsection{NMHV 6\textendash{}point amplitudes}
The first non-trivial soft-behaviour can be found when considering
NMHV helicity configurations of 6-point amplitudes.
Let us consider the amplitude
$A_{6}\left(\gamma_s^{+},1_{q}^{-},2_g^{-},3_g^{-},4_g^{+},5_{\bar{q}}^{+}\right)
\equiv
A_{6}\left(s^{+},1_{q}^{-},2^{-},3^{-},4^{+},5_{\bar{q}}^{+}\right)$,
built from BCFW recursion,
\begin{align}
A_{6}\left(s^{+},1_{q}^{-},2^{-},3^{-},4^{+},5_{\bar{q}}^{+}\right) &
=\frac{i\left\langle 35\right\rangle \left\langle 3\left|4+5\right|s\right]^{2}}
{P^2_{s12}\left\langle 34\right\rangle \left\langle 45\right\rangle \left\langle 5\left|s+1\right|2\right]\left[12\right]}
+\frac{i\left\langle 1\left|s+5\right|4\right]^{2}\left\langle 5\left|s+1\right|4\right]}
{P^2_{s15}\left\langle s1\right\rangle \left\langle s5\right\rangle \left\langle 5\left|s+1\right|2\right]\left[32\right]\left[43\right]}\, ,
\label{eq:E4}
\end{align}
with $P_{ij}=k_i+k_j$ and $P_{ijl}=k_{i}+k_{j}+k_{l}$.
We construct the soft limit, first by rescaling $|s\rangle \to \epsilon |s\rangle$,
\begin{multline}
%
A_{6}\left(s^{+},1_{q}^{-},2^{-},3^{-},4^{+},5_{\bar{q}}^{+}\right)
\overset{|s\rangle \to \epsilon |s\rangle}{=}\label{eq:E5-1}\\
=-\frac{i}{\epsilon\langle5|s|2]+\langle5|1|2]}
\left(\frac{(\epsilon\langle5|s|4]+\langle5|1|4])(\epsilon\langle1|s|4]+\langle1|5|4])^{2}}{\epsilon^{2}[3|2][4|3]\langle s|1\rangle\langle s|5\rangle\left(\epsilon(P^2_{1s}+P^2_{5s})+P^2_{15}\right)}
+\frac{\langle3|5\rangle \langle3|4+5|s]^{2}}{[2|1]\langle3|4\rangle\langle4|5\rangle\left(\epsilon(P^2_{1s}+P^2_{2s})+P^2_{12}\right)}\right)\, ,
\end{multline}
and then, by expanding around $\epsilon \to 0$, 
\bea
A_{6}\left(s^{+},1_{q}^{-},2^{-},3^{-},4^{+},5_{\bar{q}}^{+}\right)
\quad \overset{\epsilon \to 0}{=} \quad
{a_{-2} \over \epsilon^2} + 
{a_{-1} \over \epsilon} + {\cal O}(1) \ ,
\eea
with
\begin{align}
a_{-2} &
=-\frac{i[4|1][5|4]^{2}\langle1|5\rangle}{[2|1][3|2][4|3][5|1]\langle
  s|1\rangle\langle s|5\rangle} \ ,
\label{eq:SL6}\\
a_{-1} & =\frac{1}{\langle s|1\rangle}\left(-\frac{i[5|4]^{2}[4|s]}{[2|1][3|2][4|3][5|1]}+\frac{i[4|1][5|4]^{2}[2|s]}{[2|1]^{2}[3|2][4|3][5|1]}+\frac{i[4|1][5|4]^{2}[5|s]}{[2|1][3|2][4|3][5|1]^{2}}\right)\nonumber \\
 & \qquad+\frac{1}{\langle
   s|5\rangle}\left(-\frac{i[5|4][4|1]^{2}[5|s]}{[2|1][3|2][4|3][5|1]^{2}}-\frac{i[5|4][4|1][4|s]}{[2|1][3|2][4|3][5|1]}\right)
\ . \label{eq:SNL6}
\end{align}

Alternatively, the soft expansion can be obtained through the action of 
differential soft operators on the non-radiative 5-point amplitude,
according to,
\begin{eqnarray}
a_{-2} &=& S^{(0)\lambda}
A_{5}(1_{q}^{-},2^{-},3^{-},4^{+},5_{\bar{q}}^{+}) \ ,  \\
a_{-1} &=& S^{(1)\lambda} A_{5}(1_{q}^{-},2^{-},3^{-},4^{+},5_{\bar{q}}^{+}) \ ,
\end{eqnarray}
where,
the operators $S^{(0)\lambda}$ and $S^{(1)\lambda}$, respectively
defined in ~(\ref{S0_f_BCFW}) and~(\ref{S1_f_BCFW}), read
\begin{align}
S^{(0)\lambda} & =\frac{\,\langle1|5\rangle}{\langle s|1\rangle\langle
  s|5\rangle} \ , \label{S0_6p}\\
S^{(1)\lambda} & =\frac{1}{\langle
  s|1\rangle}\tilde{\lambda}_{s}^{\dot{a}}\frac{\partial}{\partial\tilde{\lambda}_{1}^{\dot{a}}}-\frac{1}{\langle
  s|5\rangle}\tilde{\lambda}_{s}^{\dot{a}}\frac{\partial}{\partial\tilde{\lambda}_{5}^{\dot{a}}}
\ ,\label{S1_6p}
\end{align}
while the 5-point quark-gluon amplitude is,
\begin{align}
A_{5}\left(1_{q}^{-},2^{-},3^{-},4^{+},5_{\bar{q}}^{+}\right) &
=\frac{i[1|4][5|4]^{3}}{[1|2][2|3][3|4][4|5][5|1]} \ . 
\end{align}
In this case the leading soft singularity is,
\begin{align}
a_{-2} = 
\frac{\langle1|5\rangle}{\langle s|1\rangle\langle s|5\rangle}A_{5}\left(1_{q}^{-},2^{-},3^{-},4^{+},5_{\bar{q}}^{+}\right)
 =-\frac{i[4|1][5|4]^{2}\langle1|5\rangle}{[2|1][3|2][4|3][5|1]\langle s|1\rangle\langle s|5\rangle}\label{eq:SL5}
\end{align}
in full agreement with the result of (\ref{eq:SL6}). \\ 
To compute the next-to-leading soft term, we combine the derivatives,
\begin{align}
\frac{1}{\langle s|1\rangle}\tilde{\lambda}_{s}^{\dot{a}}\frac{\partial}{\partial\tilde{\lambda}_{1}^{\dot{a}}}A_{5}\left(1_{q}^{-},2^{-},3^{-},4^{+},5_{\bar{q}}^{+}\right) \nonumber
&=-\frac{1}{\langle s|1\rangle}\left(\frac{[4|2] [1|s]}{[2|1] [4|1]}+\frac{[5|s]}{[5|1]}\right)A_{5}\left(1_{q}^{-},2^{-},3^{-},4^{+},5_{\bar{q}}^{+}\right)\nonumber \\
 &=\frac{1}{\langle
   s|1\rangle}\left(\frac{i[4|2][5|4]^{2}[1|s]}{[2|1]^{2}[3|2][4|3][5|1]}+\frac{i[4|1][5|4]^{2}[5|s]}{[2|1][3|2][4|3][5|1]^{2}}\right)
 \ , \label{eq:SNL5a}
\end{align}
and
\begin{align}
-\frac{1}{\langle s|5\rangle}\tilde{\lambda}_{s}^{\dot{a}}\frac{\partial}{\partial\tilde{\lambda}_{5}^{\dot{a}}}A_{5}\left(1_{q}^{-},2^{-},3^{-},4^{+},5_{\bar{q}}^{+}\right) \nonumber
& =\frac{1}{\langle s|5\rangle}\left(\frac{[4|s]}{[5|4]}+\frac{[4|1] [5|s]}{[5|1] [5|4]}\right)A_{5}\left(1_{q}^{-},2^{-},3^{-},4^{+},5_{\bar{q}}^{+}\right)\nonumber \\
 &=\frac{1}{\langle
   s|5\rangle}\left(-\frac{i[5|4][4|1]^{2}[5|s]}{[2|1][3|2][4|3][5|1]^{2}}-\frac{i[5|4][4|1][4|s]}{[2|1][3|2][4|3][5|1]}\right)
 \ , 
\end{align}
yielding,
\begin{align}
a_{-1} = 
\frac{1}{\langle s|1\rangle}\left(\frac{i[4|2][5|4]^{2}[1|s]}{[2|1]^{2}[3|2][4|3][5|1]}+\frac{i[4|1][5|4]^{2}[5|s]}{[2|1][3|2][4|3][5|1]^{2}}\right)+\frac{1}{\langle s|5\rangle}\left(-\frac{i[5|4][4|1]^{2}[5|s]}{[2|1][3|2][4|3][5|1]^{2}}-\frac{i[5|4][4|1][4|s]}{[2|1][3|2][4|3][5|1]}\right)\, .\label{eq:SNL5}
\end{align}
After applying the Schouten relation, the result in (\ref{eq:SNL6})
becomes identical to  (\ref{eq:SNL5}). \\ 

The agreement between the direct soft-limit expansion and the
application of the soft operators can be analogously verified 
for the other NMHV helicity configurations,  
$A_{6}\left(s^{+},1_{q}^{-},2^{+},3^{-},4^{-},5_{\bar{q}}^{+}\right)$
and $A_{6}\left(s^{+},1_{q}^{-},2^{-},3^{+},4^{-},5_{\bar{q}}^{+}\right)$.

\section{Soft-limit of quark-gluon amplitudes}
\label{sec:soft-gluon-QCD}

In this section, we discuss the low-energy behaviour of soft-gluon radiation from quark-gluon tree-level amplitudes.
Depending on the position of the soft-gluon $g_s$ within the colour-ordered amplitude, we may have three situations \cite{Giele:1991vf},
\begin{align}
&A_{n+3}\left(\Lambda_{q};
  g_{1},\cdots,g_{n},g_s;\Lambda_{\bar{q}}\right) \ , \label{sg-antiquark-gluon}\\
&A_{n+3}\left(\Lambda_{q};
  g_s,g_{1},\cdots,g_{n};\Lambda_{\bar{q}}\right) \ , \label{sg-quark-gluon}\\
&A_{n+3}\left(\Lambda_{q}; g_{1},\cdots,g_m,g_s,g_{m+1},\cdots,g_{n};\Lambda_{\bar{q}}\right)\,. \label{sg-gluon-gluon}
\end{align}
The first two cases can be considered specular to each other, since
they describe the soft-gluon adjacent to one fermion and one gluon,
while the third case represents the soft-gluon adjacent to two gluons. 
Therefore, we will consider as independent only the first and the
third configurations, which are discussed in the followings.
For both cases, by making use of the results in sections \ref{sec:2}
and \ref{sec:3}, we will establish the equivalence of the soft
operators derived {\it via} gauge invariance and on-shell recurrence,
representing the main result of this work.

\begin{figure}[h]
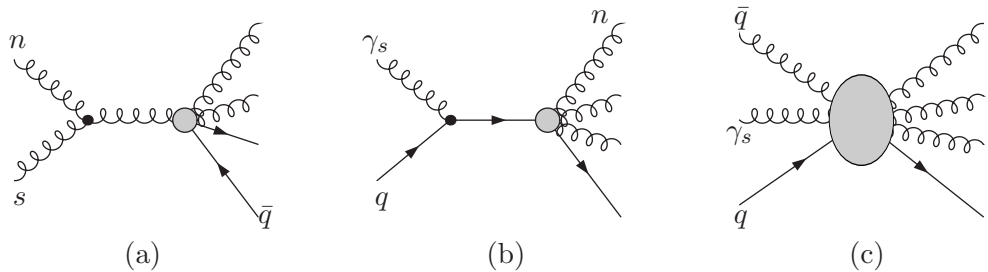

\begin{align*}
\parbox{40mm}{\input{FeynmanDiagrams/fig3a.tex}}\qquad
\parbox{40mm}{\input{FeynmanDiagrams/fig3b.tex}}\qquad
\parbox{40mm}{\input{FeynmanDiagrams/fig3c.tex}}
\end{align*}
\caption[a]{Soft-gluon behaviour of quark-gluon amplitudes: case 1}
\label{fig:gSq}
\end{figure}

\subsection{Case 1: soft-gluon adjacent to the anti-quark and one
  gluon}
\label{sec:qcd:case1}
\hspace{0pt} 
The colour-ordered amplitude $A_{n+3}\left(\Lambda_{q};
  g_{1},\cdots,g_{n},g_s;\Lambda_{\bar{q}}\right)$ 
describes the soft-gluon emitted from the external anti-quark
$\overline q$ or from the external gluon $g_{n}$ or from internal
gluon lines between $g_{n}$ and $\overline {q}$, and receives
contributions from three types of diagrams, as shown in fig.~\ref{fig:gSq}.
 
Following the procedure presented in sec.~\ref{sec:3}, from on-shell
recursion relation, we can derive the soft behaviour,
\bea
A_{n+3}\left(\Lambda_{q};
  g_{1},\cdots,g_{n},g_s;\Lambda_{\bar{q}}\right)=
\left({1\over \epsilon^2}S_{QCD}^{(0)\lambda}+{1\over
    \epsilon}S_{QCD}^{(1)\lambda}\right)A_{n+2}
\left(\Lambda_q; g_1,\cdots, g_n; \Lambda_{\bar    q}\right)
+ {\cal O}(1) \ ,
\label{NS_BCFW}
\eea
with
\bea
S_{QCD}^{(0)\lambda}&=&{\Spaa{n,\,\bar q}\over  \Spaa{\bar q,\,s} \Spaa{s,\,n} }\, ,\label{S0_QCD_BCFW}\\
S_{QCD}^{(1)\lambda}&=&{1\over \Spaa{s, \bar q}}\widetilde \lambda_s^{\dot a}{\partial\over \partial \widetilde \lambda_{\bar q}^{\dot a}}-{1\over \Spaa{s, n}}\widetilde \lambda_s^{\dot a}{\partial\over \partial \widetilde \lambda_{n}^{\dot a}}\, ,\label{S1_QCD_BCFW}
\eea 
and where $A_{n+2}$ is the non-radiative quark-gluon amplitude.  

Gauge invariance, on the other side, requires the amplitude to have
the following expression, 
\begin{align}
A_{n+3}(k_{q};k_{1},\dots,k_{n},k_{s};k_{\bar{q}})= &
\left(\frac{\varepsilon^{+}(k_s;r_s)\cdot
    k_{\bar{q}}}{\sqrt{2}k_{\bar{q}}\cdot
    k_{s}}-\frac{\varepsilon^{+}(k_s;r_s)\cdot
    k_{n}}{\sqrt{2}k_{n}\cdot k_{s}}\right)
A_{n+2}(k_{q};k_{1},\dots,k_{n};k_{\bar{q}})\nonumber \\
 &
 \quad+\frac{i\,\varepsilon_{\mu}^{+}(k_s;r_s)\,{k}_{s\nu}}{\sqrt{2}k_{q}\cdot
   k_{s}} J_{{\rm G}n}^{\mu\nu} 
\left[{\overline{u}}(k_{q})\,
\tilde{A}( k_{q};\, k_{1},\dots,k_{n};k_{\bar{q}})v(k_{\bar{q}})\right]\nonumber\\ 
 &
 \quad+\frac{i\,\varepsilon_{\mu}^{+}(k_s;r_s)\,{k}_{s\nu}}{\sqrt{2}k_{\bar{q}}\cdot
   k_{s}}{\overline{u}}(k_{q})
\tilde{A}(k_{q};\, k_{1},\dots,k_{n};k_{\bar{q}})\Sigma_{F}^{\mu\nu}v(k_{\bar{q}})\nonumber \\
 &\quad-\frac{i\,\varepsilon_{\mu}^{+}(k_s;r_s)\,
   k_{s\,\nu}}{\sqrt{2}}{\overline{u}}(k_{q})\left[\frac{L_{\bar{q}}^{\mu\nu}}{k_{\bar{q}}\cdot
     k_{s}}
\tilde{A}(k_{q};\, k_{1},\dots,k_{n};k_{\bar{q}})\right]v(k_{\bar{q}})
 \label{QCD-case1}
\end{align}
with the non-radiative amplitude being $A_{n+2}(k_s; k_{\bar q}, k_q, k_1, \dots, k_n)={\overline
  u}(k_q) \tilde{A}(k_{\bar q},\,k_q,\,k_1, \dots, k_n) v(k_{\bar q})$.
In the above expression,  $J^{\mu\nu}_{{\rm G}n}$ is the total angular
momentum for gluon $g_n$, defined in (\ref{Gauge_boson_Jmn}), while  
$L^{\mu\nu}_{\bar q}$ and $\Sigma_F^{\mu\nu}$ are respectively the
orbital and spin angular momenta of anti-quark $\overline q$, given in (\ref{totampli0}).

The equivalence of the two derivation can be established as follows.

\begin{itemize}
\item[{\bf A.}]
{\bf The leading soft singularity}

The leading soft term coming from 
$S_{QCD}^{(0)\lambda}$ in (\ref{S0_QCD_BCFW}) and 
$\left(\frac{\varepsilon^{+}(k_s;r_s)\cdot
    k_{\bar{q}}}{\sqrt{2}k_{\bar{q}}\cdot
    k_{s}}-\frac{\varepsilon^{+}(k_s;r_s)\cdot
    k_{n}}{\sqrt{2}k_{n}\cdot k_{s}}\right)$ in (\ref{QCD-case1})
agree with each other, once both are expressed in spinor variables, exactly as in (\ref{S0-QED-equivalence}).

\item[{\bf B.}]
{\bf The next-to-leading soft singularity}

The expression (\ref{QCD-case1}), obtained from gauge invariance, contains two contributions: 
\begin{enumerate}
\item the operator
$\frac{i\,\varepsilon_{\mu}^{+}(k_s;r_s)\,{k}_{s\nu}}{\sqrt{2}k_{q}\cdot
  k_{s}} J_{{\rm G}n}^{\mu\nu}$,  related to gluon $g_n$ comes from
fig.~\ref{fig:gSq} (a) and (c). 
This term is equivalent to 
$-{1\over \Spaa{s, n}}\widetilde \lambda_s^{\dot
  a}{\partial\over \partial \widetilde \lambda_{n}^{\dot a}}$ of
$S_{QCD}^{(1)\lambda}$ defined in (\ref{S1_QCD_BCFW}).
The proof comes from the equivalence in pure-gluon cases \cite{Bern:2014vva}, whose result we recall in sec.~\ref{sec:2},
eqs. (\ref{S1YM_polar1p_GI})-(\ref{S1YM_polar1m_BCFW}).

\item the operators $-\frac{i\,\varepsilon_{\mu}^{+}(k_s;r_s)\, k_{s\,\nu}}{\sqrt{2} k_{\bar{q}}\cdot k_{s}} L_{\bar{q}}^{\mu\nu}$ and $\frac{i\,\varepsilon_{\mu}^{+}(k_s;r_s)\,{k}_{s\nu}}{\sqrt{2}k_{\bar{q}}\cdot k_{s}}\Sigma_{F}^{\mu\nu}$ related to anti-quark $\overline q$ come from fig.~\ref{fig:gSq} (b) and (c) 
According to discussions in sec~\ref{subsec:connection_QCD},
eqs.(\ref{S1_BCFW_qbar_hp})-(\ref{S1_GI_qbar_hm}) and
(\ref{S1B-on-Mqbar})-(\ref{S1G-on-Mqbar_inv}), the combination of
these two terms is equivalent to ${1\over \Spaa{s, \bar q}}\widetilde \lambda_s^{\dot a}{\partial\over \partial \widetilde \lambda_{\bar q}^{\dot a}}$ in $S_{QCD}^{(1)\lambda}$ of (\ref{S1_QCD_BCFW}).
\end{enumerate}

\end{itemize}

Therefore we can consider the soft behaviour of (\ref{S1_QCD_BCFW}) 
equivalent to (\ref{QCD-case1}).

\subsection{Case 2: soft-gluon adjacent to two gluons}
\label{sec:qcd:case2}
\begin{figure}[h]
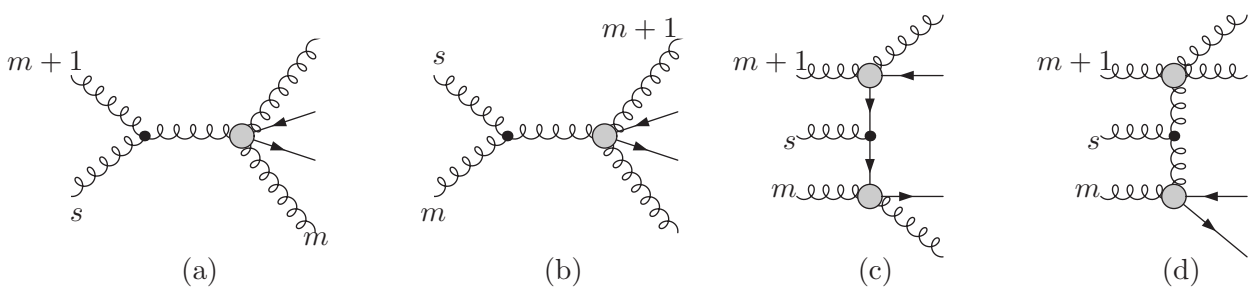

\centering
\begin{align*}
\qquad ~
\parbox{40mm}{\input{FeynmanDiagrams/fig4a.tex}}\qquad
\parbox{40mm}{\input{FeynmanDiagrams/fig4b.tex}}\qquad
\parbox{40mm}{\input{FeynmanDiagrams/fig4c.tex}}
\parbox{40mm}{\input{FeynmanDiagrams/fig4d.tex}}
\end{align*}
\caption[a]{Soft-gluon behaviour of quark-gluon amplitudes: case 2}
\label{fig:gSg}
\end{figure}

\hspace{0pt}
In the colour-ordered amplitude 
$A_{n+3}\left(\Lambda_{q};
  g_{1},\cdots,g_m,g_s,g_{m+1},\cdots,g_{n};\Lambda_{\bar{q}}\right)$,
the soft-gluon $g_s$ can be emitted from either external gluons, or
internal gluon lines between, or internal fermion line, 
as shown in fig.~\ref{fig:gSg}.

The proof of the equivalence between the derivation from the on-shell
formalisms and from gauge-invariance is identical to one of the pure-gluon
case \cite{Bern:2014vva}.
The only difference being that, in the considered case, the term fixed
by gauge invariance, called $N$ in \cite{Bern:2014vva} and coming from fig.~\ref{fig:pure-gluon} (c)
(see also sec.~\ref{subsec:QCD-gauge-inv} and fig.~\ref{GluonSoftFigure} (c), for photon emission), 
receives now contributions from two pieces, corresponding to fig.~\ref{fig:gSg} (c) and (d).

\subsection{Examples}

\subsubsection{$A_{6}\left(1_{q}^{-},2_{g}^{-},3_{g}^{-},4_{g}^{+},s_{g}^{+},5_{\bar{q}}^{+}\right)$} 

We derive the soft behaviour for a colour-ordered six-point NMHV amplitude corresponding to the case in sec.~\ref{sec:qcd:case1}.
Let us consider 
$A_{6}\left(1_{q}^{-},2_{g}^{-},3_{g}^{-},4_{g}^{+},s_{g}^{+},5_{\bar{q}}^{+}\right)\equiv A_{6}\left(1_{q}^{-},2^{-},3^{-},4^{+},s^{+},5_{\bar{q}}^{+}\right)$,
where the amplitude is written in~\cite{Luo:2005rx},
\begin{align}
A_{6}\left(1_{q}^{-},2^{-},3^{-},4^{+},s^{+},5_{\bar{q}}^{+}\right) & =-\frac{i\langle5|2+3|4]\langle1|2+3|4]^{2}}{P_{234}^{2}[3|2][4|3]\langle1|5\rangle\langle s|3+4|2]\langle s|5\rangle}-\frac{i\langle3|s+4|1]\langle3|s+4|5]^{2}}{P_{s34}^{2}[2|1][5|1]\langle3|4\rangle\langle s|3+4|2]\langle s|4\rangle}\,. 
\end{align}
We first construct the soft limit by rescaling $\left|s\right\rangle \to\epsilon\left|s\right\rangle $
\begin{multline}
A_{6}\left(1_{q}^{-},2^{-},3^{-},4^{+},s^{+},5_{\bar{q}}^{+}\right)\overset{\left|s\right\rangle \to\epsilon\left|s\right\rangle }{=}\\
=\frac{i}{\epsilon^{2}\langle s|3+4|2]}\left(\frac{(\epsilon\langle1|s|4]+\langle1|5|4])^{2}(\epsilon\langle5|s|4]+\langle5|1|4])}{[3|2][4|3]\langle1|5\rangle\langle s|5\rangle\left(\epsilon\langle s|1+5|s]+P_{15}^{2}\right)}-\frac{(\epsilon\langle3|s|1]+\langle3|4|1])(\epsilon\langle3|s|5]+\langle3|4|5])^{2}}{[2|1][5|1]\langle3|4\rangle\langle s|4\rangle\left(\epsilon\langle s|3+4|s]+P_{34}^{2}\right)}\right)\, ,
\end{multline}
and then by expanding around $\epsilon\to0$
\begin{align}
A_{6}\left(1_{q}^{-},2^{-},3^{-},4^{+},s^{+},5_{\bar{q}}^{+}\right) & =\frac{a_{-2}}{\epsilon^{2}}+\frac{a_{-1}}{\epsilon}+\mathcal{O}\left(1\right)\, ,
\end{align}
with
\begin{align}
a_{-2} & =-\frac{i[5|4]^{2}\langle3|4|1]}{[2|1][5|1]\langle s|3+4|2]\langle s|4|3]}-\frac{i[4|5|1|4][5|4]}{[3|2][4|3]\langle s|3+4|2]\langle s|5|1]}\, ,\label{eq:Ex2A2SL}\\
a_{-1} & =-\frac{i[5|4]^{2}\langle3|s|1]}{[2|1][5|1]\langle s|3+4|2]\langle s|4|3]}+\frac{i[5|4]^{2}[4|s]}{[3|2][4|3][5|1]\langle s|3+4|2]}+\frac{i[4|1][5|4]^{2}\langle s|3|s]}{[2|1][4|3][5|1]\langle s|3+4|2]\langle s|4|3]}\nonumber \\
 & \quad-\frac{i[4|1][5|4]^{2}\langle s|1|s]}{[3|2][4|3][5|1]\langle s|3+4|2]\langle s|5|1]}+\frac{i[4|1][5|4]^{2}[4|s]}{[2|1][4|3]^{2}[5|1]\langle s|3+4|2]}-\frac{i[4|1][5|4]^{2}[5|s]}{[3|2][4|3][5|1]^{2}\langle s|3+4|2]}\nonumber \\
 & \quad+\frac{2i[5|4][4|1|s|4]}{[3|2][4|3]\langle s|3+4|2]\langle s|5|1]}+\frac{2i[4|1][5|4]\langle3|s|5]}{[2|1][5|1]\langle s|3+4|2]\langle s|4|3]}\, .\label{eq:Ex2A1SL}
\end{align}

As was done in the previous example, soft expansion is obtained by
the action of differential soft operators on the non-radiative 5-point
amplitude as 
\begin{align}
a_{-2} & =S^{\left(0\right)\lambda}A_{5}\left(1_{q}^{-},2^{-},3^{-},4^{+},5_{\bar{q}}^{+}\right)\, ,\\
a_{-1} & =S^{\left(1\right)\lambda}A_{5}\left(1_{q}^{-},2^{-},3^{-},4^{+},5_{\bar{q}}^{+}\right)\,,
\end{align}
where operators $S^{\left(0\right)\lambda}$ and $S^{\left(1\right)\lambda}$
read
\begin{align}
S^{\left(0\right)\lambda} & =\frac{\langle4|5\rangle}{\langle s|4\rangle\langle s|5\rangle}\, ,\\
S^{\left(1\right)\lambda} & =\frac{1}{\langle s|5\rangle}\tilde{\lambda}_{s}^{\dot{a}}\frac{\partial}{\partial\tilde{\lambda}_{5}^{\dot{a}}}-\frac{1}{\langle s|4\rangle}\tilde{\lambda}_{s}^{\dot{a}}\frac{\partial}{\partial\tilde{\lambda}_{4}^{\dot{a}}}\, .
\end{align}
The leading soft singularity is
\begin{align}
a_{-2} & =\frac{\langle4|5\rangle}{\langle s|4\rangle\langle s|5\rangle}A_{5}\left(1_{q}^{-},2^{-},3^{-},4^{+},5_{\bar{q}}^{+}\right)=\frac{i[4|1][5|4]^{2}\langle4|5\rangle}{[2|1][3|2][4|3][5|1]\langle s|4\rangle\langle s|5\rangle}\, ,\label{eq:Ex2A2SLA}
\end{align}
and to compute the next-to-leading order soft term, we combine the
derivatives
\begin{align}
\frac{1}{\langle s|5\rangle}\tilde{\lambda}_{s}^{\dot{a}}\frac{\partial}{\partial\tilde{\lambda}_{5}^{\dot{a}}}A_{5}\left(1_{q}^{-},2^{-},3^{-},4^{+},5_{\bar{q}}^{+}\right) & =\frac{1}{\langle s|5\rangle}\left(\frac{[4|1][5|s]}{[5|1][5|4]}-\frac{[4|s]}{[5|4]}\right)A_{5}\left(1_{q}^{-},2^{-},3^{-},4^{+},5_{\bar{q}}^{+}\right)\nonumber \\
 & =\frac{1}{\langle s|5\rangle}\left(\frac{i[4|1][5|4][4|s]}{[2|1][3|2][4|3][5|1]}-\frac{i[4|1]^{2}[5|4][5|s]}{[2|1][3|2][4|3][5|1]^{2}}\right),
\end{align}
and
\begin{align}
-\frac{1}{\langle s|4\rangle}\tilde{\lambda}_{s}^{\dot{a}}\frac{\partial}{\partial\tilde{\lambda}_{4}^{\dot{a}}}A_{5}\left(1_{q}^{-},2^{-},3^{-},4^{+},5_{\bar{q}}^{+}\right) & =-\frac{1}{\langle s|4\rangle}\left(\frac{[1|s]}{[1|4]}-\frac{[3|s]}{[3|4]}+\frac{2[5|s]}{[5|4]}\right)A_{5}\left(1_{q}^{-},2^{-},3^{-},4^{+},5_{\bar{q}}^{+}\right)\nonumber \\
 & =\frac{1}{\langle s|4\rangle}\left(-\frac{i[5|4]^{2}[1|s]}{[2|1][3|2][4|3][5|1]}+\frac{i[4|1][5|4]^{2}[3|s]}{[2|1][3|2][4|3]^{2}[5|1]}+\frac{2i[4|1][5|4][5|s]}{[2|1][3|2][4|3][5|1]}\right)
\end{align}
yielding 
\begin{align}
a_{-1} & =\frac{1}{\langle s|5\rangle}\left(\frac{i[4|1][5|4][4|s]}{[2|1][3|2][4|3][5|1]}-\frac{i[4|1]^{2}[5|4][5|s]}{[2|1][3|2][4|3][5|1]^{2}}\right)\nonumber \\
 & \qquad+\frac{1}{\langle s|4\rangle}\left(-\frac{i[5|4]^{2}[1|s]}{[2|1][3|2][4|3][5|1]}+\frac{i[4|1][5|4]^{2}[3|s]}{[2|1][3|2][4|3]^{2}[5|1]}+\frac{2i[4|1][5|4][5|s]}{[2|1][3|2][4|3][5|1]}\right)\, .\label{eq:Ex2A1SLA}
\end{align}
Eqs. (\ref{eq:Ex2A2SL}) and (\ref{eq:Ex2A1SL}) agree numerically
with (\ref{eq:Ex2A2SLA}) and (\ref{eq:Ex2A1SLA}) respectively.

\subsubsection{$A_{6}\left(1_{q}^{-},2_{g}^{-},3_{g}^{-},s_{g}^{+},4_{g}^{+},5_{\bar{q}}^{+}\right)$}

We derive the soft behaviour for a color ordered six-point NMHV amplitude corresponding to the case in sec.~\ref{sec:qcd:case2}.
Let us consider
$A_{6}\left(1_{q}^{-},2_{g}^{-},3_{g}^{-},s_{g}^{+},4_{g}^{+},5_{\bar{q}}^{+}\right)\equiv A_{6}\left(1_{q}^{-},2^{-},3^{-},s^{+},4^{+},5_{\bar{q}}^{+}\right)$,
where the amplitude is written in~\cite{Luo:2005rx},
\begin{align*}
A_{6}\left(1_{q}^{-},2^{-},3^{-},s^{+},4^{+},5_{\bar{q}}^{+}\right) & =\frac{i\langle1|2+3|s]^{2}\langle5|2+3|s]}{P_{s23}^{2}[3|2]\langle1|5\rangle\langle4|5\rangle[3|s]\langle4|s+3|2]}-\frac{i\langle3|s+4|1]\langle3|s+4|5]^{2}}{P_{s34}^{2}[2|1][5|1]\langle4|s+3|2]\langle s|3\rangle\langle s|4\rangle}\, .
\end{align*}
We first construct the soft limit by rescaling $\left|s\right\rangle \to\epsilon\left|s\right\rangle $
\begin{multline}
A_{6}\left(1_{q}^{-},2^{-},3^{-},s^{+},4^{+},5_{\bar{q}}^{+}\right)\overset{\left|s\right\rangle \to\epsilon\left|s\right\rangle }{=}\\
=\frac{i}{\epsilon\langle4|s|2]+\langle4|3|2]}\left(\frac{\langle1|2+3|s]^{2}\langle5|2+3|s]}{[3|2]\langle1|5\rangle\langle4|5\rangle[3|s]\left(\epsilon\langle s|2+3|s]+P_{23}^{2}\right)}-\frac{(\epsilon\langle3|s|1]+\langle3|4|1])(\epsilon\langle3|s|5]+\langle3|4|5])^{2}}{\epsilon^{2}[2|1][5|1]\langle s|3\rangle\langle s|4\rangle\left(\epsilon\langle s|3+4|s]+P_{34}^{2}\right)}\right)\, ,
\end{multline}
and then by expanding around $\epsilon\to0$
\begin{align}
A_{6}\left(1_{q}^{-},2^{-},3^{-},s^{+},4^{+},5_{\bar{q}}^{+}\right) & =\frac{a_{-2}}{\epsilon^{2}}+\frac{a_{-1}}{\epsilon}+\mathcal{O}\left(1\right)
\end{align}
with
\begin{align}
a_{-2} & =\frac{i[4|1][5|4]^{2}\langle3|4\rangle}{[2|1][3|2][4|3][5|1]\langle s|3\rangle\langle s|4\rangle}\, ,\label{eq:Ex3A2SL}\\
a_{-1} & =\frac{1}{\langle s|3\rangle}\left(\frac{i[4|1][5|4]^{2}[2|s]}{[2|1][3|2]^{2}[4|3][5|1]}-\frac{i[4|1][5|4]^{2}[4|s]}{[2|1][3|2][4|3]^{2}[5|1]}\right)\nonumber \\
 & \quad+\frac{1}{\langle s|4\rangle}\left(\frac{i[5|4]^{2}[1|s]}{[2|1][3|2][4|3][5|1]}-\frac{i[4|1][5|4]^{2}[3|s]}{[2|1][3|2][4|3]^{2}[5|1]}-\frac{2i[4|1][5|4][5|s]}{[2|1][3|2][4|3][5|1]}\right)\, .\label{eq:Ex3A1SL}
\end{align}
The soft expansion is obtained by the action of differential soft
operators on the non-radiative 5-point amplitude as 
\begin{align}
a_{-2} & =S^{\left(0\right)\lambda}A_{5}\left(1_{q}^{-},2^{-},3^{-},4^{+},5_{\bar{q}}^{+}\right)\, ,\\
a_{-1} & =S^{\left(1\right)\lambda}A_{5}\left(1_{q}^{-},2^{-},3^{-},4^{+},5_{\bar{q}}^{+}\right)\, ,
\end{align}
where operators $S^{\left(0\right)\lambda}$ and $S^{\left(1\right)\lambda}$
read
\begin{align}
S^{\left(0\right)\lambda} & =\frac{\langle4|3\rangle}{\langle s|4\rangle\langle s|3\rangle}\, ,\\
S^{\left(1\right)\lambda} & =\frac{1}{\langle s|4\rangle}\tilde{\lambda}_{s}^{\dot{a}}\frac{\partial}{\partial\tilde{\lambda}_{4}^{\dot{a}}}-\frac{1}{\langle s|3\rangle}\tilde{\lambda}_{s}^{\dot{a}}\frac{\partial}{\partial\tilde{\lambda}_{3}^{\dot{a}}}\, .
\end{align}
The leading soft singularity is
\begin{align}
a_{-2} & =\frac{\langle3|4\rangle}{\langle s|4\rangle\langle3|s\rangle}A_{5}\left(1_{q}^{-},2^{-},3^{-},4^{+},5_{\bar{q}}^{+}\right)=\frac{i[4|1][5|4]^{2}\langle4|5\rangle}{[2|1][3|2][4|3][5|1]\langle s|4\rangle\langle s|5\rangle}\, ,\label{eq:Ex3A2SLA}
\end{align}
which agrees with the result of (\ref{eq:Ex3A2SL}).\\
For the next-to-leading order soft term, we compute the derivatives
\begin{align}
\frac{1}{\langle s|4\rangle}\tilde{\lambda}_{s}^{\dot{a}}\frac{\partial}{\partial\tilde{\lambda}_{4}^{\dot{a}}}
A_{5}\left(1_{q}^{-},2^{-},3^{-},4^{+},5_{\bar{q}}^{+}\right) 
& =\frac{1}{\langle s|4\rangle}\left(\frac{[1|s]}{[1|4]}-\frac{[3|s]}{[3|4]}+\frac{2[5|s]}{[5|4]}\right)
A_{5}\left(1_{q}^{-},2^{-},3^{-},4^{+},5_{\bar{q}}^{+}\right)\nonumber \\
 & =\frac{1}{\langle s|4\rangle}\left(\frac{i[5|4]^{2}[1|s]}{[2|1][3|2][4|3][5|1]}-\frac{i[4|1][5|4]^{2}[3|s]}{[2|1][3|2][4|3]^{2}[5|1]}-\frac{2i[4|1][5|4][5|s]}{[2|1][3|2][4|3][5|1]}\right)\\
-\frac{1}{\langle s|3\rangle}\tilde{\lambda}_{s}^{\dot{a}}\frac{\partial}{\partial\tilde{\lambda}_{3}^{\dot{a}}}
A_{5}\left(1_{q}^{-},2^{-},3^{-},4^{+},5_{\bar{q}}^{+}\right) 
& =-\frac{1}{\langle s|3\rangle}\left(-\frac{[2|s]}{[2|3]}-\frac{[4|s]}{[4|3]}\right)
A_{5}\left(1_{q}^{-},2^{-},3^{-},4^{+},5_{\bar{q}}^{+}\right)\nonumber \\
 & =\frac{1}{\langle s|3\rangle}\left(\frac{i[4|1][5|4]^{2}[2|s]}{[2|1][3|2]^{2}[4|3][5|1]}-\frac{i[4|1][5|4]^{2}[4|s]}{[2|1][3|2][4|3]^{2}[5|1]}\right)
\end{align}
yielding
\begin{align}
a_{-1} & =\frac{1}{\langle s|3\rangle}\left(\frac{i[4|1][5|4]^{2}[2|s]}{[2|1][3|2]^{2}[4|3][5|1]}-\frac{i[4|1][5|4]^{2}[4|s]}{[2|1][3|2][4|3]^{2}[5|1]}\right)\nonumber \\
 & \quad+\frac{1}{\langle s|4\rangle}\left(\frac{i[5|4]^{2}[1|s]}{[2|1][3|2][4|3][5|1]}-\frac{i[4|1][5|4]^{2}[3|s]}{[2|1][3|2][4|3]^{2}[5|1]}-\frac{2i[4|1][5|4][5|s]}{[2|1][3|2][4|3][5|1]}\right)
\end{align}
in full agreement with (\ref{eq:Ex3A1SL}).

\section{Conclusion}
\label{sec:conclusion}

We have shown that the low-energy behaviour of radiative tree-level amplitudes in QCD, 
when the momentum of the emitted particle becomes soft, 
is governed by the non-radiative process and depends on the quantum data of the emitter. 
The low-energy expansion is captured by universal operators, 
whose form is dictated by gauge invariance.
While the well known leading soft term is expressed as an eikonal factor, 
the subleading soft term depends on the total angular momentum of the emitter.
We have shown that, within the spinor formalism, remarkably, 
the subleading soft operator of single-gluon emission from quark-gluon amplitudes appears as a differential operator whose form does not depend on the spin of the emitter.
Our result, derived from gauge invariance and on-shell recursive construction, is, therefore, 
in line with the results recently derived for pure gluon- and graviton-scattering.

\section*{Acknowledgement}
We thank Zvi Bern for stimulating discussions, and Mingxing Luo, Valery Yundin and
Gionata Luisoni for clarifying considerations.
H.L. is supported by the ERC Advanced Grant no. 267985 (DaMeSyFla). 
The work of P.M. is supported by the Alexander von Humboldt
Foundation, in the framework of the Sofja Kovalevskaja Award 2010,
endowed by the German Federal Ministry of Education and Research.
W.J.T. is supported by Fondazione Cassa di Risparmio di Padova e Rovigo (CARIPARO).
\appendix
\section{Low-energy regular terms}
\label{sec:Appendix}
By following similar arguments in \cite{Cachazo:2014fwa},
we can show that the third term in (\ref{eq:bcfw:qqg})
is regular in the soft limit. 

The terms contributing to the regular part of the soft expansions in (\ref{eq:bcfw:qqg}) come from poles of $A_{n+1}(z)/z$ located at solutions of 
\bea
\left(k_s(z)+k_q+k_{\bar q}+k_{a_1}+\cdots +k_{a_m}\right)^2 =0 \label{z-sqqbargluon-eq}
\eea
with $m\geq 0$ ($m=0$ means no non-soft gluon on the same side of soft-gluon). 
Denote $Q_{q\bar q m}=k_q+k_{\bar q}+k_{a_1}+\cdots +k_{a_m}$. The solution to (\ref{z-sqqbargluon-eq}) is
\bea
z^*=-{(k_s+Q_{q\bar q m})^2\over 2 p\cdot Q_{q\bar q m}}
\eea 
With $p=|n\rangle |n]$ and $p\cdot k_s =0$. 
The terms contributing to non-3 point structures are
\bea
\sum_{h_I=\pm 1}A_{L}\left(\Lambda_{\bar q}, \widehat{\gamma_s^+},\Lambda_q,g_1,\ldots,g_m,\widehat{I} \right)\frac{1}{(k_s+Q_{q\bar q m})^2}A_{R}\left(-\widehat{I},g_{m+1},\ldots,\widehat{g_n}\right)
\eea
While taking limit $|s\rangle \rightarrow 0$, one can treat $z^*$ value as
\begin{alignat}{2}
z^*=-{Q_{q\bar q m}^2\over 2 p\cdot Q_{q\bar q m}} \quad k_I= Q_{q \bar q m}-{Q_{q\bar q m}^2\over 2 p\cdot Q_{q \bar q m}} p
\end{alignat}
$k_I$ is the projection of $Q_m$ along the null direction defined by $p$ and null as well.

The left amplitude in the limit $|s\rangle \rightarrow 0$ is 
\bea
A_{L}\left(\Lambda_{\bar q}, \widehat{\gamma_s^+},\Lambda_q,g_1,\ldots,g_m,\widehat{I} \right)\rightarrow A_{L}\left(\Lambda_{\bar q}, z^* p,\Lambda_q,g_1,\ldots,g_m,\widehat{I} \right)
\eea
This amplitude with $z^* p$ momenta in the soft limit contains no singularities, i.e., soft, collinear or multi-particle singularities, thus is all finite.  Since $Q_{q \bar q m}\neq 0$, which is not the case for singularity part in the limit.

On the other hand,  the right amplitude under the limit is
\bea
A_{R}\left(-\widehat{I},g_{m+1},\ldots, g_{n-1},\widehat{g_n}\right)\rightarrow A_{R}\left(-\widehat{I},g_{m+1},\ldots,g_{n-1}, k_n(z^*)\right)
\eea
with $k_n(z^*)=k_n-z^* p$. The only possible dangerous term is remaining single gluon except $g_n$, say $m+1=n-1$ above. Then the right amplitude becomes 
\bea
A_{R}\left(\{k_I, -h_I\},\{k_{n-1},h_{n-1}\}, \{k_n(z^*),h_n\}\right)
\eea
From momentum conservation $k_s+Q_{q\bar q m}+k_{n-1}+k_n$, we have $Q_{q\bar q m}=-k_s-k_{n-1}-k_n$ and moreover, under the limit $k_s\rightarrow 0$, we get $Q_{q\bar q m}=-k_{n-1}-k_n$. 
In this case, we can do the computation from the deformation in right amplitude
\bea
k_I&=&\left(-|n-1\rangle -{\Spbb{s,n}\over \Spbb{s,n-1}}|n\rangle \right)|n-1],\\
k_n(z^*)&=&{\Spbb{s,n}\over \Spbb{s,n-1}}|n\rangle|n-1]
\eea
Notice all fermion lines are in the left amplitude, the right amplitude contains only pure-gluon interactions.
If the right amplitude includes two positive helicities and one negative helicity, it vanished directly; if the right amplitude includes two negative herlicities and one positive helicity, the contribution is finite by taking only the holomorphic component of the soft limit $|s\rangle \rightarrow 0$. 

We can, therefore, conclude that
singularities only appear in the first two terms of
(\ref{eq:bcfw:qqg}), as considered in sec. \ref{sec:3}.


\bibliographystyle{jhep}
\bibliography{References_soft}

\end{document}